\documentclass[3p,times,procedia]{elsarticle}
\usepackage{nupha_ecrc}

\volume{00}

\firstpage{1}

\journalname{Nuclear Physics A}

\runauth{}

\jid{nupha}

\jnltitlelogo{Nuclear Physics A}

\usepackage{amssymb}
\usepackage{lineno}

\usepackage[figuresright]{rotating}

\begin{document}

\begin{frontmatter}

%% Title, authors and addresses

%% use the tnoteref command within \title for footnotes;
%% use the tnotetext command for the associated footnote;
%% use the fnref command within \author or \address for footnotes;
%% use the fntext command for the associated footnote;
%% use the corref command within \author for corresponding author footnotes;
%% use the cortext command for the associated footnote;
%% use the ead command for the email address,
%% and the form \ead[url] for the home page:
%%
%% \title{Title\tnoteref{label1}}
%% \tnotetext[label1]{}
%% \author{Name\corref{cor1}\fnref{label2}}
%% \ead{email address}
%% \ead[url]{home page}
%% \fntext[label2]{}
%% \cortext[cor1]{}
%% \address{Address\fnref{label3}}
%% \fntext[label3]{}

%% Instructions from Editor: Please use the following \dochead only in the preprint version (e-print arXiv etc.); 
%% use empty \dochead{} when submitting to Nuclear Physics A!
\dochead{XXVIIIth International Conference on Ultrarelativistic Nucleus-Nucleus Collisions\\ (Quark Matter 2019)}
%\dochead{}
%% Use \dochead if there is an article header, e.g. \dochead{Short communication}
%% \dochead can also be used to include a conference title, if directed by the editors
%% e.g. \dochead{17th International Conference on Dynamical Processes in Excited States of Solids}

\title{Z production in pPb collisions and charmonium production in PbPb ultra-peripheral collisions at LHCb}

%% use optional labels to link authors explicitly to addresses:
%% \author[label1,label2]{<author name>}
%% \address[label1]{<address>}
%% \address[label2]{<address>}

\author[label1]{Hengne Li}
\author[]{on behalf of the LHCb collaboration}

\address[label1]{Guangdong Provincial Key Laboratory of Nuclear Science, Institute of Quantum Matter, South China Normal University, Guangzhou 510006, China.}

\begin{abstract}
%% Text of abstract
This article presents recent results  
of the Z boson production in the proton-lead collisions and the 
charmonium production in the ultra-peripheral lead-lead collisions at the LHC 
collected during 2013 to 2018 by the LHCb detector. 
The potential heavy ion programs in these topics are also discussed. 

\end{abstract}

\begin{keyword}
%% keywords here, in the form: keyword \sep keyword
Heavy-ion collisions, nuclear modification, Z boson production,
ultra-peripheral collisions, charmonium production, LHCb experiment,
LHC 
%% MSC codes here, in the form: \MSC code \sep code
%% or \MSC[2008] code \sep code (2000 is the default)

\end{keyword}

\end{frontmatter}

%%
%% Start line numbering here if you want
%%
%\linenumbers

%% main text
\section{Introduction}
\label{sec:introduction}

The LHCb detector~\cite{LHCb-DP-2008-001, LHCb-DP-2014-002} 
is a fully instrumented single-arm spectrometer in the forward region 
covering a pseudorapidity acceptance of
$2 < \eta < 5$,  providing a high tracking momentum resolution down to very low 
transverse momentum ($p_{\rm T}$) and precise vertex reconstruction capability. 
The detector is originally designed for heavy-flavour measurements, precision tests of the standard model (SM),
and searches for physics beyond the SM.
As a young member of the LHC heavy ion program, 
the LHCb experiment also provides the unique datasets for 
the heavy ion physics studies at LHC. 
%The detector has a pseudorapidity acceptance of
%$2 < \eta < 5$,  providing a high tracking momentum resolution down to very low 
%transverse momentum ($p_{\rm T}$) thanks to
%the high-precision tracking system consisting of a silicon-strip vertex 
%detector surrounding the beam-beam interaction region, a 
%large-area silicon-strip detector located upstream of a dipole magnet with a bending power of about 
%4\,Tm, and three stations of silicon-strip detectors and straw drift tubes placed downstream of the magnet.
%Different types of charged hardons are distinguished using information from 
%two ring-imaging Cherenkov detectors. Photons, electrons and hadrons are identified
%by a calorimeter system consisting of scintillating-pad and preshower detectors,
% and an electromagnetic and a hadronic calorimeter. 
% Muons are identified by a system composed of alternating layers of iron and 
% multiwire proportional chambers.
The heavy ion program at LHCb includes proton-lead (pPb), lead-lead (PbPb),
and xenon-xenon (XeXe) collisions, together with 
fixed-target collisions of p or Pb with noble gases nuclei
%atoms
%, such as 
%helium (He), neon (Ne), argon (Ar), 
injected 
%in a range of about one meter 
around the interaction point. 
The summary of the LHCb heavy ion data in collision mode
and recorded integrated luminosities are given in Table~\ref{tab:dataset}.
%The kinematic acceptance of the data sets taken by the LHCb detector 
%in terms centre-of-mass energy square (${\rm s_{NN}}$) versus rapidity 
%in the centre-of-mass frame ($y^*$) is presented in Figure~\ref{fig:kin_acc}.

In this article, we present recent results  
of the Z boson production in the proton-lead collisions~\cite{Aaij:2014pvu, LHCb:CONF2019003} and the 
charmonium production in the ultra-peripheral lead-lead collisions~\cite{LHCb-CONF-2018-003} at the LHC 
using datasets collected during 2013 to 2018 by the LHCb detector. 

\begin{table}[htbp]
\begin{center}
\begin{footnotesize}
\begin{tabular}{c|cccc|ccc}
                       & \multicolumn{2}{c}{2013}       &   \multicolumn{2}{c|}{2016}     & 2015  &   2017 & 2018 \\  
$\sqrt{s_{\rm NN}}$ &  \multicolumn{2}{c}{5.02\,TeV}  &  \multicolumn{2}{c|}{8.16\,TeV} & 5.02\,TeV  & 5.02\,TeV  & 5.02\,TeV \\ 
                        \hline
                        & pPb     &  Pbp         &   pPb   &  Pbp    & PbPb   & XeXe   &   PbPb \\

$\mathcal{L}$ & 1.1\,nb$^{-1}$ & 0.5\,nb$^{-1}$ &13.6\,nb$^{-1}$ &20.8\,nb$^{-1}$ &10\,$\mu$b$^{-1}$ &0.4\,$\mu$b$^{-1}$ &$\sim$210\,$\mu$b$^{-1}$ \\
\end{tabular}
\end{footnotesize}
\end{center}
\vspace*{-0.5cm}
\caption{Summary of the LHCb heavy ion collisions and the recorded integrated luminosities.}
\label{tab:dataset}
\end{table}

%\begin{figure}[htbp]
%\begin{center}
% \includegraphics[width=0.5\linewidth]{kin_accept.pdf}
%\caption{Summary of the kinematic acceptance of the data sets taken by the LHCb detector in terms centre-of-mass energy square (${\rm s_{NN}}$) versus rapidity in the centre-of-mass frame ($y^*$ ).}
%\label{fig:kin_acc}
%\end{center}
%\end{figure}

\section{Z boson production in proton-lead collisions}
\label{sec:zproduction}

Electroweak bosons are unmodified by the hot and dense 
medium created in heavy ion collisions.
Their leptonic decays pass through the medium 
without being affected by strong interactions.
Therefore, electroweak boson productions well retained 
the initial conditions of the collisions, can be
used to probe nuclear matter effects and constraint nuclear parton distribution functions
(nPDFs) for Bjorken-x from $\sim10^{-4}$ to 1 at Q$^2 \sim 10^4$\,GeV$^2$~\cite{Kusina:2016fxy,Eskola:2016oht},
and can be used as a calibration of the nuclear 
modification factor of other processes.

Using pPb datasets collected by the LHCb detector at 
5.02\,TeV and 8.16\,TeV as summarized in Table~\ref{tab:dataset},
the Z boson production cross-sections in the muon decay channel are measured in the 
fiducial volume in both the forward (p-Pb) and backward (Pb-p) collision 
configurations~\cite{Aaij:2014pvu, LHCb:CONF2019003} 
based on the following equation:
$\sigma_{{\rm Z}\to\mu^+\mu^-} = \left[{\rm N_{cand}}\cdot \rho\right]/\left[\mathcal{L}\cdot \epsilon\right]$,
where $\sigma_{{\rm Z}\to\mu^+\mu^-}$ is the production cross-section to be measured,
${\rm N_{cand}}$ is the number of $ {\rm Z}\to\mu^+\mu^-$ candidates pass 
signal selection, 
$\rho$ is the signal purity of the selected Z candidates,
$\mathcal{L}$ is the integrated luminosity, 
and $\epsilon$ is the total efficiency
includes trigger, reconstruction and selection efficiencies. 
The fiducial volume is defined as $60<m_{\mu^+\mu^-}<120$\,GeV,
$2.0<\eta_{\mu^{\pm}}<4.5$, and $p_{\rm T}^{\mu^{\pm}}>20$\,GeV.
The purity is measured using data-driven methods, and the efficiencies are estimated 
using Monte-Carlo (MC) samples together with tag-and-probe data driven corrections.

\begin{figure}[h]
%  \begin{center}
%    \includegraphics[width=0.49\linewidth]{ppb_mass_5tev.pdf}
%    \put(-170,70){(a)}
%    \includegraphics[width=0.49\linewidth]{pbp_mass_5tev.pdf}
%    \put(-170,70){(b)}
%    \vspace*{-0.5cm}
%  \end{center}
  \begin{center}
    \includegraphics[width=0.33\linewidth]{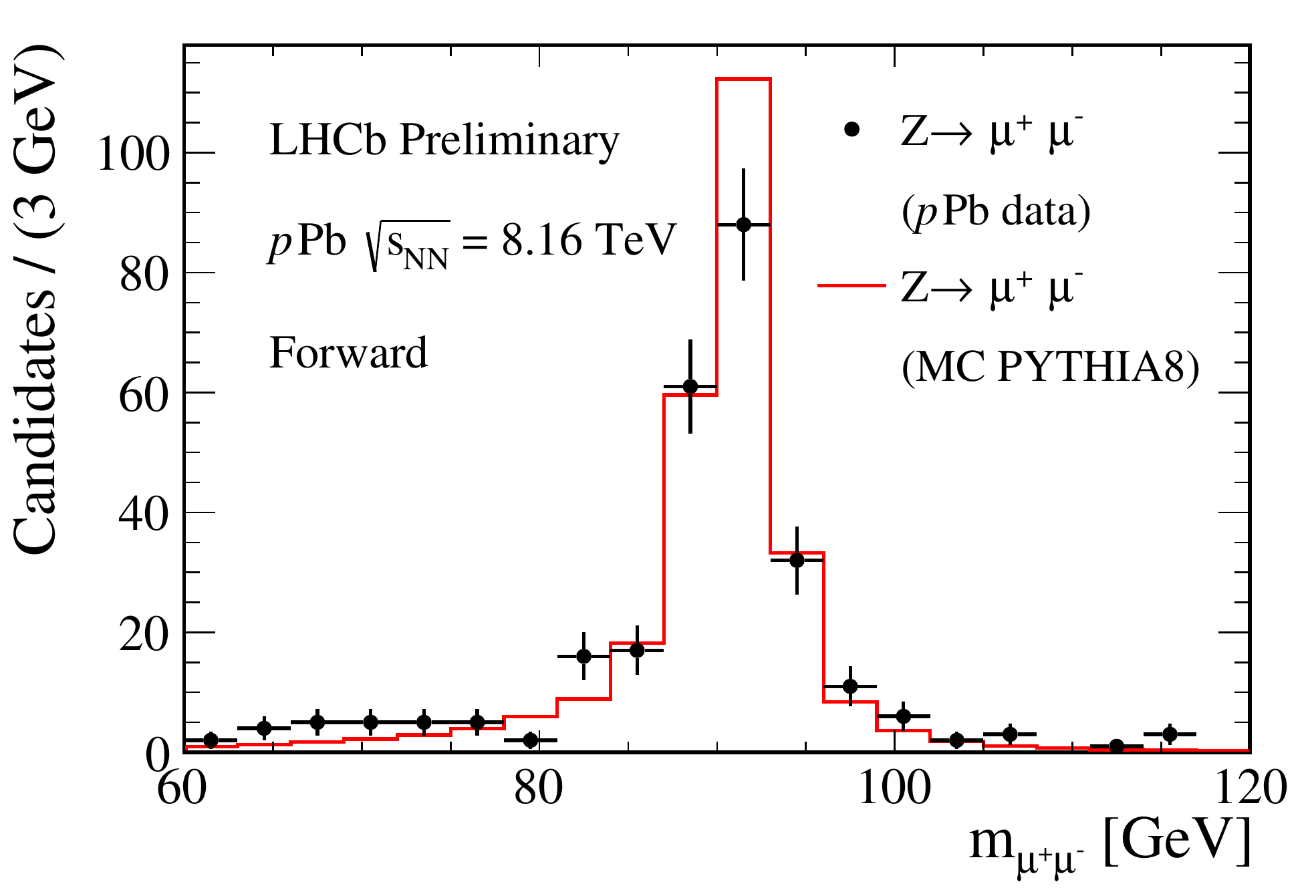}
    \put(-110,30){(a)}
    \includegraphics[width=0.33\linewidth]{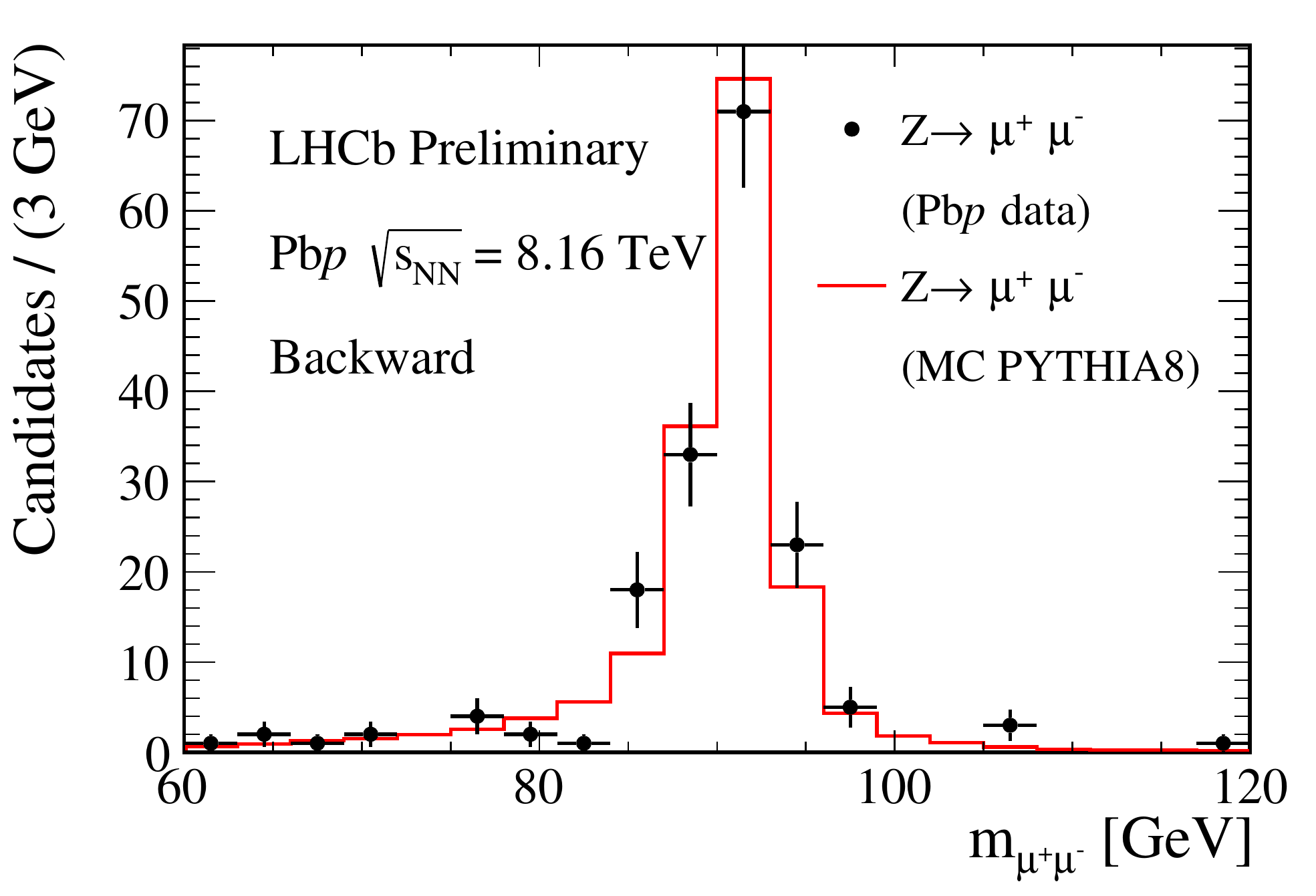}
    \put(-110,30){(b)}
    \vspace*{-0.8cm}
  \end{center}
  \caption{(color online) The dimuon invariant mass distributions after the offline selection
for pPb (a) and Pbp (b) configurations,
using datasets taken 
%in 2013 at $\sqrt{s_{\rm NN}} = 5.02$\,TeV 
%(a, b) and 
in 2016 at $\sqrt{s_{\rm NN}} = 8.16$\,TeV.
The red line shows the distributions from 
simulation generated using PYTHIA 8~\cite{Sjostrand:2007gs} with CTEQ6L1~\cite{Stump:2003yu} PDF set, 
normalised to the number of observed candidates. }
  \label{fig:candidates}
\end{figure}

The invariant mass distributions of selected signal candidates
are shown in Fig.~\ref{fig:candidates} for datasets 
taken in 2016 at $\sqrt{s_{\rm NN}} = 8.16$\,TeV.

\begin{figure}[h]
\begin{center}
 \includegraphics[width=0.32\linewidth]{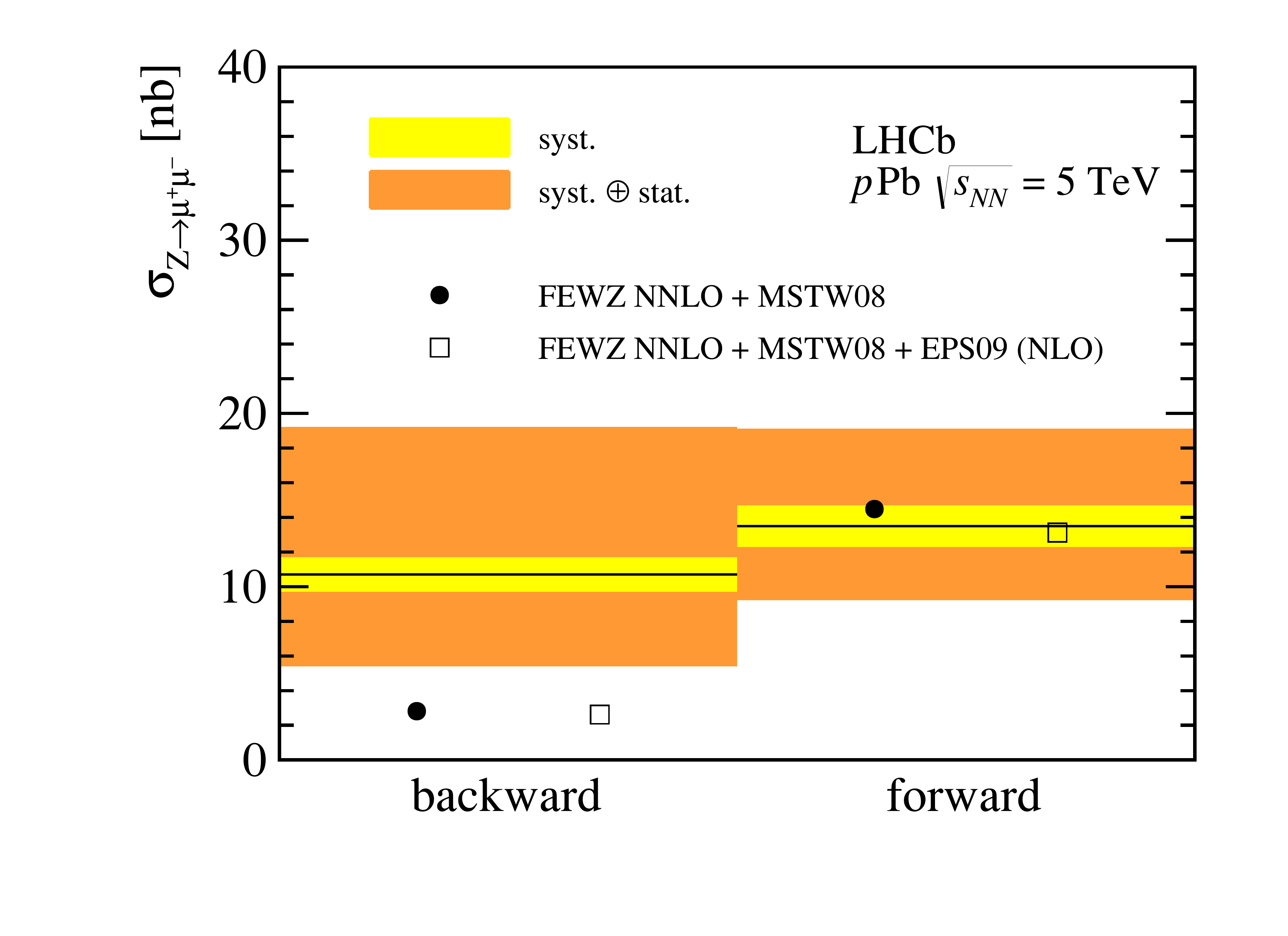}
     \put(-60,23){(a)}
 \includegraphics[width=0.33\linewidth]{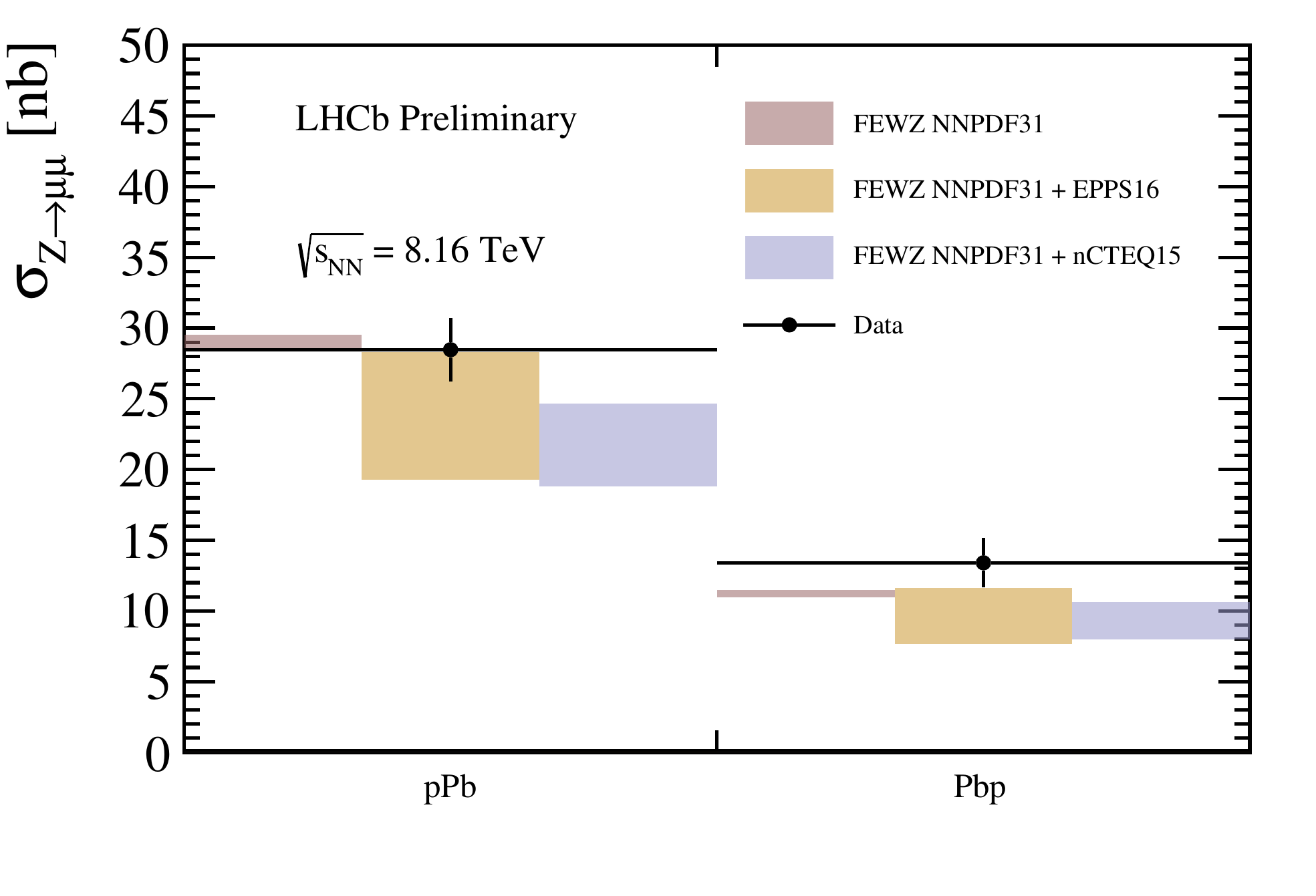}
     \put(-70,23){(b)}
  \includegraphics[width=0.33\linewidth]{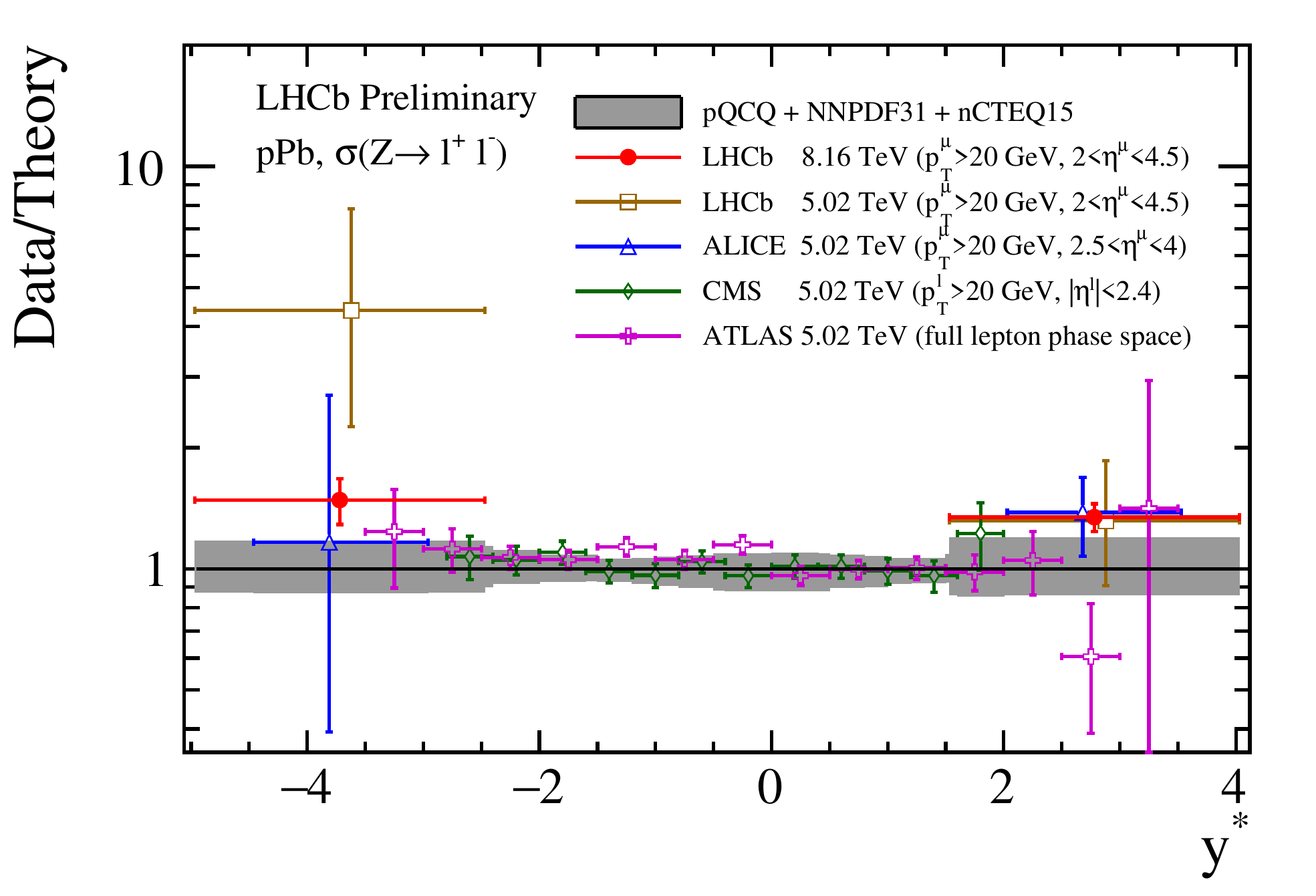}
     \put(-70,23){(c)}
\vspace*{-0.8cm}
\end{center}
\caption{(color online) (a, b) Measured fiducial cross-sections of Z production
and theoretical calculations using various PDF sets with or without nuclear modification.
Figure (a) is for dataset taken in 2013 at $\sqrt{s_{\rm NN}} = 5.02$\,TeV  
and figure (b) is for dataset taken in 2016 at $\sqrt{s_{\rm NN}} = 8.16$\,TeV.
(c) Comparison of LHCb 8.16\,TeV results with previous 5.02\,TeV results from ATLAS, 
CMS, ALICE, and LHCb. 
The uncertainties on the data over theory ratios include only the experimental statistical and systematic uncertainties; 
the PDF uncertainties are shown separately on the line at one. 
The central values of the LHCb and ALICE results at 5.02\,TeV are shifted to left and right by 
0.1 units in rapidity, respectively, for better visibility.}
  \label{fig:result_cross_section}
\end{figure}

The resulting fiducial cross-sections for centre-of-mass energies 
at 5.02\,TeV and 8.16\,TeV~\cite{Aaij:2014pvu, LHCb:CONF2019003} 
are shown in Fig.~\ref{fig:result_cross_section} (a) and (b), respectively.
The results are compared with theoretical calculations using
FEWZ~\cite{Gavin:2010az,Li:2012wna} with MSTW08~\cite{Martin:2009iq} free nucleon PDF together EPS09 (NLO)~\cite{Eskola:2009uj} nPDF for dataset at 5.02\,TeV,
and with NNPDF 3.1~\cite{Ball:2017nwa} free nucleon PDF together EPPS16 (NLO)~\cite{Eskola:2016oht} 
 and nCTEQ15 (NLO)~\cite{Kusina:2016fxy, Kovarik:2015cma} nPDFs
for dataset at 8.16\,TeV. 
The results are also compared with previous 5.02\,TeV results from 
LHCb~\cite{Aaij:2014pvu}, ATLAS~\cite{Aad:2015gta},
CMS~\cite{Khachatryan:2015pzs}, and ALICE~\cite{Alice:2016wka}, 
as shown in Fig.~\ref{fig:result_cross_section} (c).
The new LHCb 8.16\,TeV results are compatible with 
previous 5.02\,TeV results at LHCb, but with about 20 times higher statistics,
it is the most precise measurement at forward region at LHC.

The ratio of the Z boson production cross-sections 
for forward and backward configurations, $R_\mathrm{FB}$, 
is particularly sensitive to nuclear effects. 
$R_\mathrm{FB}$ is measured in the common 
rapidity region ($2.5<|y^*|<4.0$) in the
centre-of-mass frame of the produced Z boson 
using 2016 dataset~\cite{LHCb:CONF2019003}  as
%\begin{equation}
    $R_\mathrm{FB}^{2.5<|y^*|<4.0}= 1.28\pm0.14({\rm stat})\pm0.14({\rm syst})\pm0.05({\rm lumi})$,
%\end{equation}
which is compatible with theoretical calculations using FEWZ with the following nPDFs:
%\begin{eqnarray*}
 $R_\mathrm{FB,NNPDF3.1+EPPS16}^{2.5<|y^*|<4.0} = 1.45 \pm 0.10{\rm (theo.)} \pm 0.01 {\rm (num.)} \pm 0.27 {\rm (nPDF)}$, and 
 $R_\mathrm{FB,NNPDF3.1+nCTEQ15}^{2.5<|y^*|<4.0} = 1.44 \pm 0.10{\rm (theo.)} \pm 0.01 {\rm (num.)} \pm 0.20 {\rm (nPDF)}$,
%\end{eqnarray*}
where, the uncertainty num. from the numerical precision.

%\begin{figure}[htbp]
%\begin{center}
% \includegraphics[width=0.45\linewidth]{rlhcb2013.pdf}
% \vspace*{-0.5cm}
%\end{center}
%\caption{(color online) Comparison of LHCb 8.16\,TeV results with previous 5.02\,TeV results from ATLAS, 
%CMS, ALICE, and LHCb. 
%The uncertainties on the data over theory ratios include only the experimental statistical and systematic uncertainties; 
%the PDF uncertainties are shown separately on the line at one. 
%The central values of the LHCb and ALICE results at 5.02\,TeV are shifted to left and right by 
%0.1 units in rapidity, respectively, for better visibility. }
%  \label{fig:compare_xsec}
%\end{figure}

\section{Charmonium production in ultra-peripheral lead-lead collisions}
\label{sec:upc}

Ultra-peripheral collisions (UPC)~\cite{BALTZ20081,doi:10.1146/annurev.nucl.55.090704.151526} 
refer to the photon-induced interactions
when two nuclei 
bypass each other with an impact parameter 
larger than the sum of their radii.
In case of nuclei with higher atomic numbers such as lead (Pb),
the photon-induced interactions are enhanced by the strong electromagnetic field,
which include photon-photon and 
photon-pomeron interactions. 
%The photon-photon interaction
%can be calculated precisely with perturbative electroweak theory.
Charmonium can be produced
in the photon-pomeron interactions, where
a photon emitted from one nucleus converted to a $c\bar{c}$ pair
interacts with a pair of gluon exchanges (pomeron) emitted
from the other nucleus.  
These productions can be distinguished as (1) coherent if the photon 
interact coherently with the whole nucleus, or (2) incoherent if the photon interacts with a single nucleon 
inside the nucleus. 
%The former case leaves the nucleus intact, while the 
%later case breaks the nucleus apart due to the relatively larger transverse momentum exchange 
%with one particular nucleon inside the nucleus.
Because of the gluon exchange mechanism in the UPC charmonium production,
such production is expected to probe the nuclear gluon distribution functions for 
Bjorken-$x$ between $\sim 10^{-5}$ and $10^{-2}$ with a hard scales of ${\rm Q^{2} \sim m^2_{J/\psi,\psi(2S)}/4}$. 
The UPC charmonium productions are therefore essential to constrain 
the uncertainty over the initial state, which is currently 
limiting the measurements of certain fundamental properties 
of the quark gluon plasma, such as the viscosity, to a high precision.
%~\cite{Akiba:2015jwa}. 

\begin{figure}[h]
\begin{center}
 \includegraphics[width=0.5\linewidth]{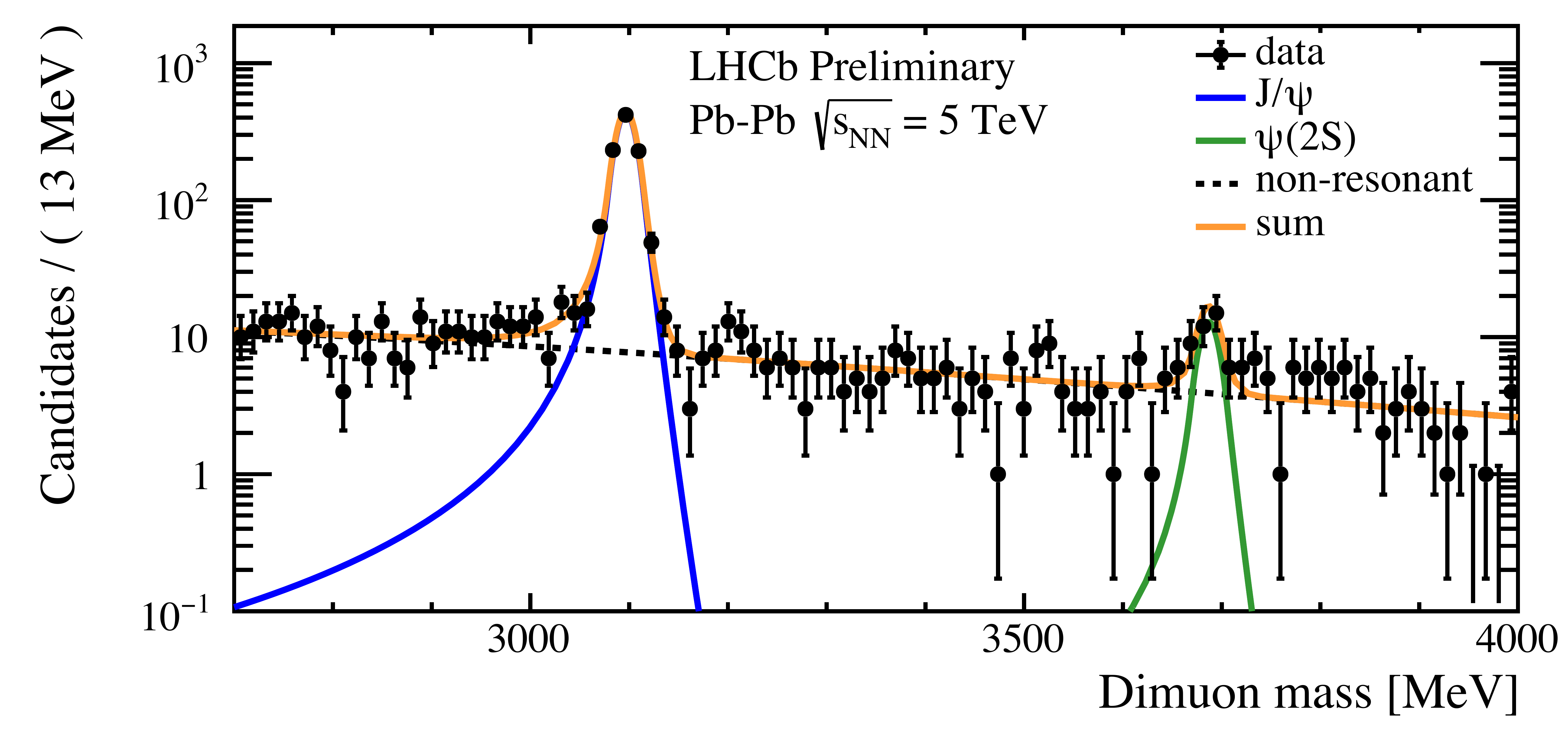}
      \put(-172,85){(a)}
 \includegraphics[width=0.33\linewidth]{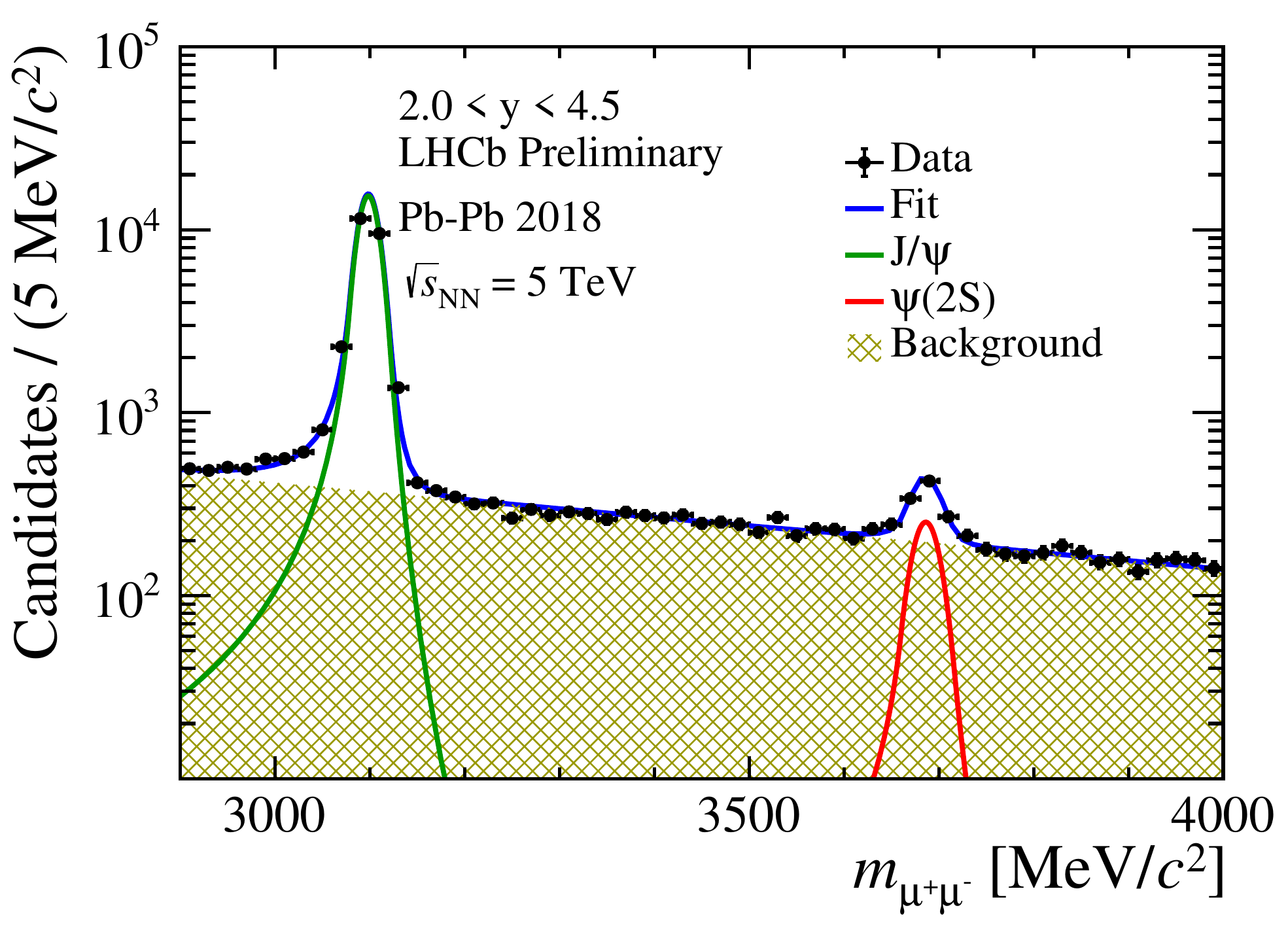}
      \put(-115,85){(b)}
\vspace*{-0.8cm}
\end{center}
\caption{(color online) Mass distributions fit to determine the fractions 
of $J/\psi$ and $\psi({\rm 2S})$ and non-resonant events. 
Figure (a) is for 2015 dataset, and figure (b) is for 2018 dataset.
 }
  \label{fig:upc_mass}
\end{figure}

\begin{figure}[h]
\begin{center}
 \includegraphics[width=0.5\linewidth]{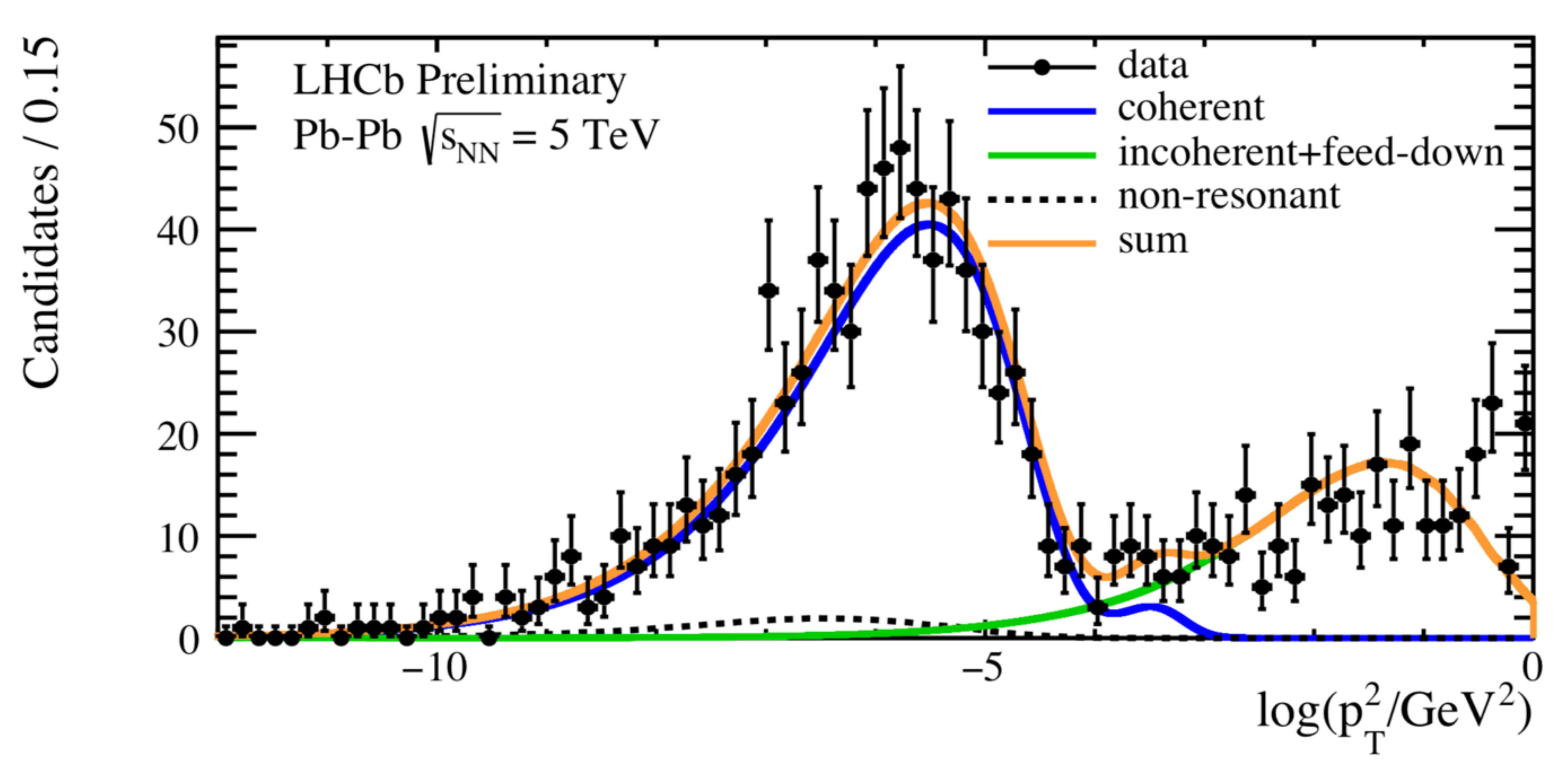}
      \put(-175,30){(a)}
  \includegraphics[width=0.34\linewidth]{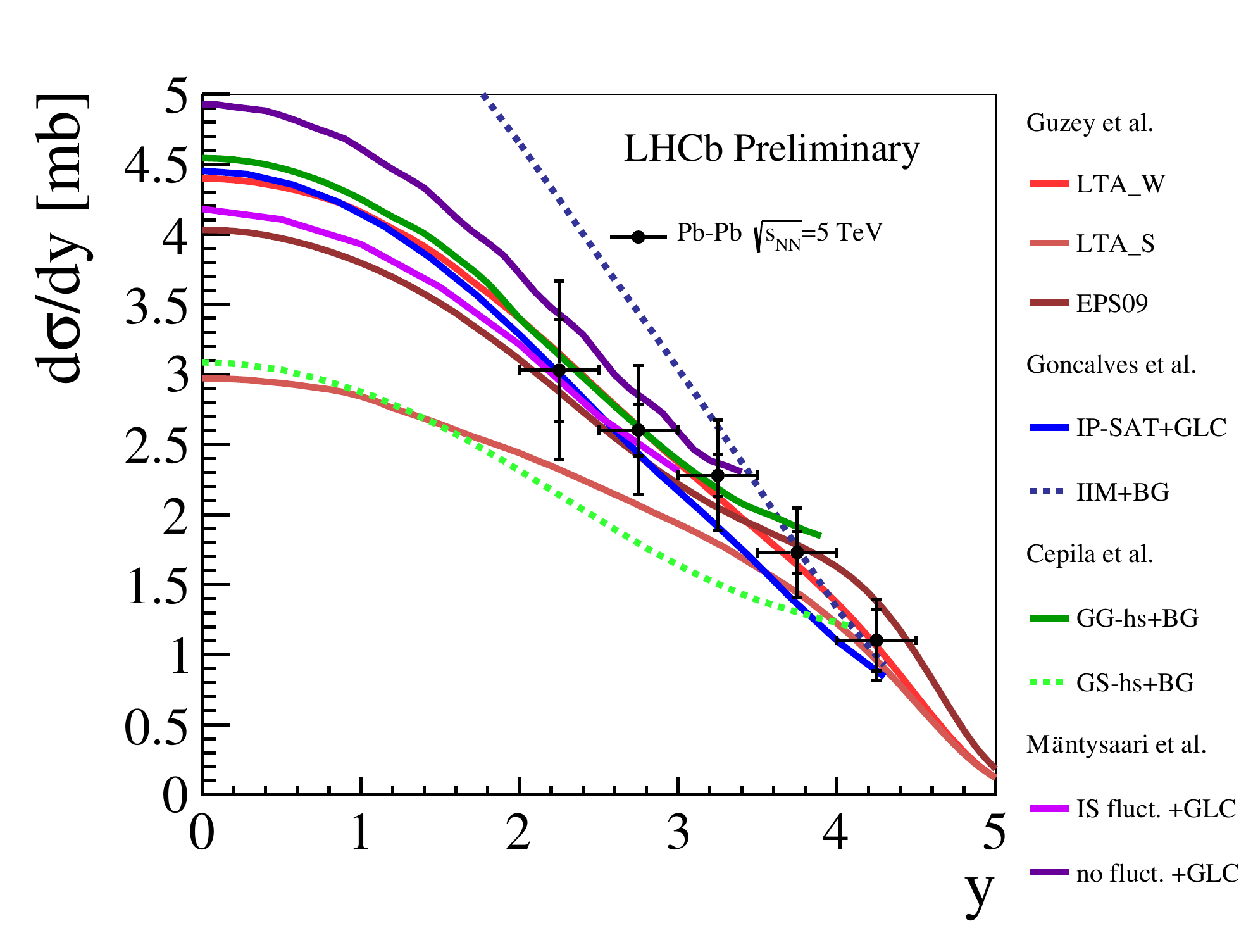}
       \put(-115,30){(b)}
\vspace*{-0.8cm}
\end{center}
\caption{(color online) (a) Fit to the distribution of ${\rm log(p_T^2)}$ of dimuon candidates after all requirements have been applied to extract the yields of coherent
and incoherent $J/\psi$ productions. 
(b) Differential cross-section for coherent $J/\psi$ production measured using 2015 PbPb dataset compared to 
different phenomenological predictions.  }
  \label{fig:upc_results_2015}
\end{figure}

Considering the extremely low transverse momentum exchange,
the event selection of the UPC charmonium production requires
a near empty detector with two long muon tracks reconstructed.
The inclusive $J/\psi$ and $\psi({\rm 2S})$ production yields are
extracted from the dimuon invariant mass fit, as shown in 
Fig.~\ref{fig:upc_mass}\,(a) for the LHCb 2015 PbPb dataset
corresponding to an integrated luminosity of about 10\,$\mu$b$^{-1}$~\cite{LHCb-CONF-2018-003},
and (b) for the 2018 PbPb dataset corresponding integrated luminosity about 20 times higher than the 2015 dataset.
The coherent and incoherent parts are then distinguished using 
a fit to the ${\rm \log(p_T^2)}$ spectrum, as shown in 
Fig.~\ref{fig:upc_results_2015}\,(a) for the 2015 dataset.
The same  ${\rm \log(p_T^2)}$ fit is performed for five separated
charmonium rapidity bins to extract the coherent $J/\psi$ production yields 
differentially.
The resulting coherent $J/\psi$ production differential cross-section measured using the 2015 data sample
is shown in Fig.~\ref{fig:upc_results_2015}\,(b),
which is compatible with most of the theoretical predictions shown in the figure,
including pQCD calculations~\cite{Guzey2016piu} and 
parametrisation based on the framework of colour-dipole 
model~\cite{Goncalves:2017wgg,PhysRevC.97.024901,MANTYSAARI2017832}.  

\section{Conclusion and outlook}
\label{sec:conclusion}

LHCb provides a unique opportunity to probe the cold nuclear matter effects using Z boson production.
Results of pPb collisions at 5.02\,TeV and 8.16\,TeV results are presented,
which are compatible with theoretical predictions involving nPDFs, where
8.16\,TeV results give the highest precision in the forward region at LHC.
Charmonium production in ultra-peripheral PbPb collisions are of particular interests to probe gluon PDFs.
Results based on 2015 PbPb dataset at 5.02\,TeV are presented.
Higher precision results using 2018 dataset with 20 times higher statistics for both the $J/\psi$ and $\psi({\rm 2S})$ cross-sections are coming soon.

%% The Appendices part is started with the command \appendix;
%% appendix sections are then done as normal sections
%% \appendix

%% \section{}
%% \label{}

%% References
%%
%% Following citation commands can be used in the body text:
%% Usage of \cite is as follows:
%%   \cite{key}         ==>>  [#]
%%   \cite[chap. 2]{key} ==>> [#, chap. 2]
%%

%% References with BibTeX database:

\bibliographystyle{elsarticle-num}
\bibliography{main}

%% Authors are advised to use a BibTeX database file for their reference list.
%% The provided style file  formats references in the required Procedia style

%% For references without a BibTeX database:

% \begin{thebibliography}{00}

%% \bibitem must have the following form:
%%   \bibitem{key}...
%%

% \bibitem{}

% \end{thebibliography}

\end{document}